\documentclass[useAMS,usenatbib]{mn2e}

\usepackage{natbib}
\usepackage{graphicx}
\usepackage{amssymb}

\title[AGN feedback and triggering of star formation]{AGN feedback and triggering of star formation in galaxies}
\author[W. Ishibashi $\&$ A. C. Fabian]
{W. Ishibashi\thanks{E-mail:
wako@ast.cam.ac.uk} 
and A. C. Fabian 
\footnotemark[0]\\
Institute of Astronomy, Madingley Road, Cambridge CB3 0HA}

\begin{document}

\pdfminorversion=4

\date{Accepted ? Received ?; in original form ? }

\pagerange{\pageref{firstpage}--\pageref{lastpage}} \pubyear{2012}

\maketitle

\label{firstpage}

\begin{abstract}
Feedback from the central black hole in active galactic nuclei (AGN) may be responsible for establishing the observed $M_\mathrm{BH} - \sigma$ relation and limiting the bulge stellar mass of the host galaxy. 
Here we explore the possibility of AGN feedback triggering star formation in the host galaxy. 
We consider a shell of dusty gas, driven outwards by radiation pressure, and analyse its escape/trapping condition in the galactic halo for different underlying dark matter potentials.Ê
In the isothermal potential, we obtain that the standard condition setting the observed $M_\mathrm{BH} - \sigma$ relation is not sufficient to clear gas out of the entire galaxy; whereas the same condition is formally sufficient in the case of the Hernquist and Navarro-Frenk-White profiles. 
The squeezing and compression of the inhomogeneous interstellar medium during the ejection process can trigger star formation within the feedback-driven shell. We estimate the resulting star formation rate and total additional stellar mass. 
In this picture, new stars are formed at increasingly larger radii and successively populate the outer regions of the host galaxy. 
This characteristic pattern may be compared with the observed `inside-out' growth of massive galaxies.  
\end{abstract}



\section{Introduction}

Black holes in active galactic nuclei (AGN) respond to the accretion process by feeding back energy to the surroundings in both radiative and kinetic forms. 
The interaction between AGN feedback and the ambient material can have a great impact on scales of the host galaxy and beyond. 
The energy released by accretion onto the central black hole clearly exceeds the binding energy of the host galaxy bulge, and it is now widely agreed that feedback from the central AGN plays an important role in the formation and evolution of the host galaxy. However, the question is how this energy and momentum resulting from the accretion process is coupled and interacts with the surrounding medium. This can in principle be achieved in a variety of ways, involving both mechanical and radiative processes, such as jets, winds, and radiation pressure.  

Observations indicate a number of direct relationships between the mass of the central black hole and the characteristic properties of the host galaxy. 
One of the tightest relations is given by the empirical correlation observed between the black hole mass and the bulge stellar velocity dispersion, of the form $M_\mathrm{BH} \propto \sigma^4$, known as the $M_\mathrm{BH} - \sigma$ relation \citep{Gebhardt_et_2000,Ferrarese_Merritt_2000}. 
A similar relation holds between the black hole mass and the bulge stellar mass, leading to $M_\mathrm{BH} \sim 10^{-3} M_\mathrm{bulge}$ \citep{Marconi_Hunt_2003, Haring_Rix_2004}.  
The existence of such correlations suggests a close coupling between the evolution of the central black hole and that of its host galaxy. 

The observational correlations can be interpreted in terms of AGN feedback, and different feedback models have been proposed and discussed in the literature \citep{Silk_Rees_1998, Fabian_1999, Fabian_et_2002, King_2003, King_2005, Murray_et_2005, Fabian_et_2006, Fabian_et_2008, King_2010, Silk_Nusser_2010, King_et_2011}. 
A simple derivation of the $M_\mathrm{BH} - \sigma$ relation is obtained by considering momentum balance, which leads to a scaling of the form $M_\mathrm{BH} = \frac{f_\mathrm{g} \sigma_\mathrm{T}}{\pi G^2 m_\mathrm{p}} \sigma^4 \, (= M_\mathrm{\sigma})$, where $f_\mathrm{g}$ is the gas mass fraction and $\sigma_\mathrm{T}$ is the Thomson cross section \citep[e.g.][and references therein]{Fabian_2012}. 
This is in remarkable agreement with the observational relation.
It is generally assumed that once the central black hole reaches the critical mass, $M_\mathrm{\sigma}$, the resulting feedback blows away the surrounding material from the galaxy bulge, eventually clearing gas out of the host galaxy. 
The removal of the ambient material, which forms the potential fuel for the feeding of the central black hole and the gas reservoir for star formation, prevents further accretion and formation of stars. 
This leads to a switch-off of the black hole growth and a quenching of star formation in the host galaxy, as seen in some numerical simulations of galaxy evolution \citep[e.g.][]{Springel_et_2005, DiMatteo_et_2005}. 
Thus the black hole seems to self-regulate its own growth, and at the same time limit the stellar mass content of the host by expelling surrounding material.  
According to this standard view, the connection between the central black hole and its host galaxy is essentially via a negative feedback mechanism.  

However, feedback from the central black hole might play additional roles in the evolution of its host galaxy.  
Here we explore the possibility of AGN feedback triggering star formation in the host galaxy \citep{Fabian_2012}. 
The interstellar medium of a galaxy is composed by cold and dense molecular gas, which form potential sites for star formation. 
Radiation from the central black hole is expected to be absorbed by dust embedded in the gas. 
We consider a shell of dusty gas being swept outwards by radiation pressure, and follow its evolution in the galactic potential. 
We assume that all of the radiation momentum output is available for driving the shell, and we ignore the details of radiative transfer. 
The squeezing and compression of the inhomogeneous interstellar medium may trigger star formation within the outflowing shell.
Stars that are formed will eventually be dropped out of the expanding shell, and start to follow their own orbits in the gravitational potential of the host galaxy.  
Several such episodes of star formation, due to AGN feedback activity, may be envisaged during the lifetime of a galaxy. 
At each episode, mass is deposited at outer radii with a fraction of gas being converted into stars.
This process adds stellar mass in the outer regions of the galaxy, thus contributing to the growth of the host. 
In order to test the plausibility of such a mechanism, we start by examining more in detail the behaviour of the feedback-driven shell in different gravitational potentials due to the contributions of dark matter and old stars.

The paper is organized as follows. 
We first discuss different physical mechanisms possibly driving AGN feedback (Sect. \ref{Sect_Wind_RadPress}). 
We next examine the escape/trapping conditions of the outflowing shell in different underlying dark matter potentials (Sect. \ref{Sect_escape_conditions}). 
In Sect. \ref{Sect_SF} we explicitly introduce star formation in the expanding shell, and estimate the star formation rate along with the corresponding additional stellar mass. 
The resulting implications are discussed in Sect. \ref{Sect_Discussion}.


\section{Winds and radiation pressure}
\label{Sect_Wind_RadPress}

As mentioned in the Introduction, AGN feedback can be driven by different physical mechanisms. 
One way of driving large-scale outflows from the central nucleus is via high-velocity winds \citep{King_2003, King_2005}.
Indeed there is observational evidence of winds and outflows detected in a number of objects \citep{Pounds_et_2003, Sturm_et_2011, Tombesi_et_2012}. 
The wind is assumed to be launched from the immediate vicinity of the central black hole, with an outflow momentum of the same order of the Eddington limited photon momentum: $\dot M_w v \sim L_E/c$. 
As the wind is generated in the small central region, the mass flow rate is limited and the outflow velocity must be very high. 
Based on the assumption that the wind mass flow rate is equal to the Eddington mass accretion rate ($\dot M_w \sim \dot M_E$),  one obtains that $v/c \sim \eta$, where $\eta \sim 0.1$ is the standard radiative efficiency \citep{King_2010, Zubovas_King_2012}. 
The derived high velocity ($v \sim 0.1 c$) corresponds to the escape speed at a launch radius of a few hundred gravitational radii. When such high-velocity winds encounter the cold interstellar medium of the host galaxy, strong shocks will inevitably result. The hot shocked gas is cooled via inverse Compton emission; but the cooling process is only efficient in the inner regions, where the cooling time is shorter than the flow time. 
Beyond a critical radius (of the order of $\sim 1$ kiloparsec) there is no way to efficiently cool the shocked gas, and the initially momentum-driven flow, considered further by \citet{McQuillin_McLaughlin_2012}, becomes energy-driven. 
The subsequent evolution of the energy-driven outflow is extensively analysed in \citet{King_2010} and \citet {King_et_2011}.  

 We do not further pursue this issue here, and next consider the direct effects of radiation pressure on dusty gas, as first introduced for AGN feedback by \citet{Fabian_1999} and discussed in \citet{Fabian_et_2002}.  
In this case, the main interaction is attributed to absorption of the central radiation by dust grains embedded in the gas. 
The radiative momentum is transferred to the gas through the coupling between dust and gas, which is further enhanced in the central environment close to the ionising source where both dust and gas are partially charged \citep{Murray_et_2005, Fabian_et_2008}. 
This strong coupling ensures that gas can be efficiently dragged out with the dust. 
We note that sublimation of dust is relevant only within the inner $\sim$parsec region (on scales smaller than the broad line region of the central AGN, cf Murray et al. 2005) and should not be an issue on the scales discussed here. 
We assume that the dusty gas which is being driven by radiation pressure is strongly radiating and remains cold ($T \ll10^4$ K), while outer gas swept up by the expanding shell is shocked to high temperatures ($\sim 10^6 - 10^7$ K) and subsequently cools down. 
Provided that a significant amount of the dust survives sputtering and other destructive processes in the shocks \citep[for a discussion on dust grain survival, see][]{Jones_Nuth_2011}, the mass of swept up material in the shell steadily increases.  
A critical radius can be defined, beyond which the outflowing shell becomes optically thin to the central radiation. 
The exact location of the critical radius depends on the coupling process, i.e. electron scattering or dust absorption. 
It is well known that the dust cross section ($\sigma_d$) is much larger than the Thomson cross section ($\sigma_T$), and the ratio is found to be of the order of $\sigma_d/\sigma_T \sim 1000$ \citep[][and references therein]{Fabian_2012}.
Dust opacities are typically of several hundred $\mathrm{cm^2/g}$ \citep[][]{Murray_et_2005}, and the corresponding critical radius is of the order of a few tens of kiloparsecs. 
Thus radiation pressure on dust grains is another viable process of driving gas out to large radii, and in the following we consider the dynamics of flows driven by radiation pressure on dusty gas.


\section{Escape/trapping conditions}
\label{Sect_escape_conditions}

We assume that radiation pressure on dusty gas sweeps up the surrounding material into a shell, and follow its temporal evolution in the gravitational potential of the host galaxy. 
We start by examining the conditions for the escape (or trapping) of the outflowing shell from the galactic halo, which is a prerequisite for the further study of star formation triggered within the feedback-driven shell.
The general form of the equation of motion of the shell is given by: 
\begin{equation}
\frac{d}{dt} [M_\mathrm{g}(r) \dot{r}] = \frac{L}{c} - \frac{G M_\mathrm{g}(r) M_\mathrm{DM}(r)}{r^2} \, ,  
\label{Eq_motion}
\end{equation} 
where $L$ is the luminosity of the central source, $M_\mathrm{g}(r)$ the enclosed gas mass, and $M_\mathrm{DM}(r)$ the dark matter mass. 
We assume that at sufficiently large radii ($r \gg G M_\mathrm{BH}/\sigma^2$), the black hole contribution can be neglected such that the potential is dominated by the dark matter component.   
In the following, we consider the evolution of the expanding shell in three types of underlying dark matter potentials, and derive the shell escape/trapping conditions in a direct and simple scheme. A more complex and detailed analysis of the dynamics of momentum-driven shells can be found in \citet{McQuillin_McLaughlin_2012}.


\subsection{Isothermal potential}

Modelling the galaxy as an isothermal sphere, the density profile of dark matter is given by
\begin{equation}
\rho(r) = \frac{\sigma^2}{2 \pi G r^2} \, , 
\end{equation}
where $\sigma$ is the velocity dispersion. 
The enclosed mass inside a sphere of radius $r$ is
\begin{equation}
M_\mathrm{DM}(r) = 4 \pi \int \rho(r) r^2 dr = \frac{2 \sigma^2}{G} r \, , 
\end{equation}
The gas mass is assumed to be a fraction $f_\mathrm{g}$ of the dark matter mass
\begin{equation}
M_\mathrm{g}(r) = f_\mathrm{g} M_\mathrm{DM}(r) = \frac{2 f_\mathrm{g} \sigma^2}{G} r \, . 
\end{equation} 
In the isothermal case, an analytical solution to the equation of motion can be derived. Integrating twice we obtain: 
\begin{equation}
r(t) = \sqrt{r_0^2 + 2 r_0 v_0 t + \left( \frac{G L}{2 f_\mathrm{g} c \sigma^2} - 2 \sigma^2 \right) t^2} \, , 
\end{equation} 
where $r_0$ and $v_0$ are the initial position and velocity, respectively. 
We note that this solution is analogous to the result previously obtained by \citet{King_2010} for the shock pattern far from the central black hole. 
The velocity is given by $v(t) = \frac{dr}{dt}$. 
The velocity of the shell can also be expressed as a function of radius as:
\begin{equation}
v(r) = \sqrt{ \frac{r_0^2 v_0^2}{r^2} +  \left( \frac{G L}{2 f_\mathrm{g} c \sigma^2} - 2 \sigma^2 \right) \left( 1 - \frac{r_0^2}{r^2} \right)} \, . 
\label{Eq_vr}
\end{equation} 

\begin{figure}
\begin{center}
\includegraphics[angle=0,width=0.35\textwidth]{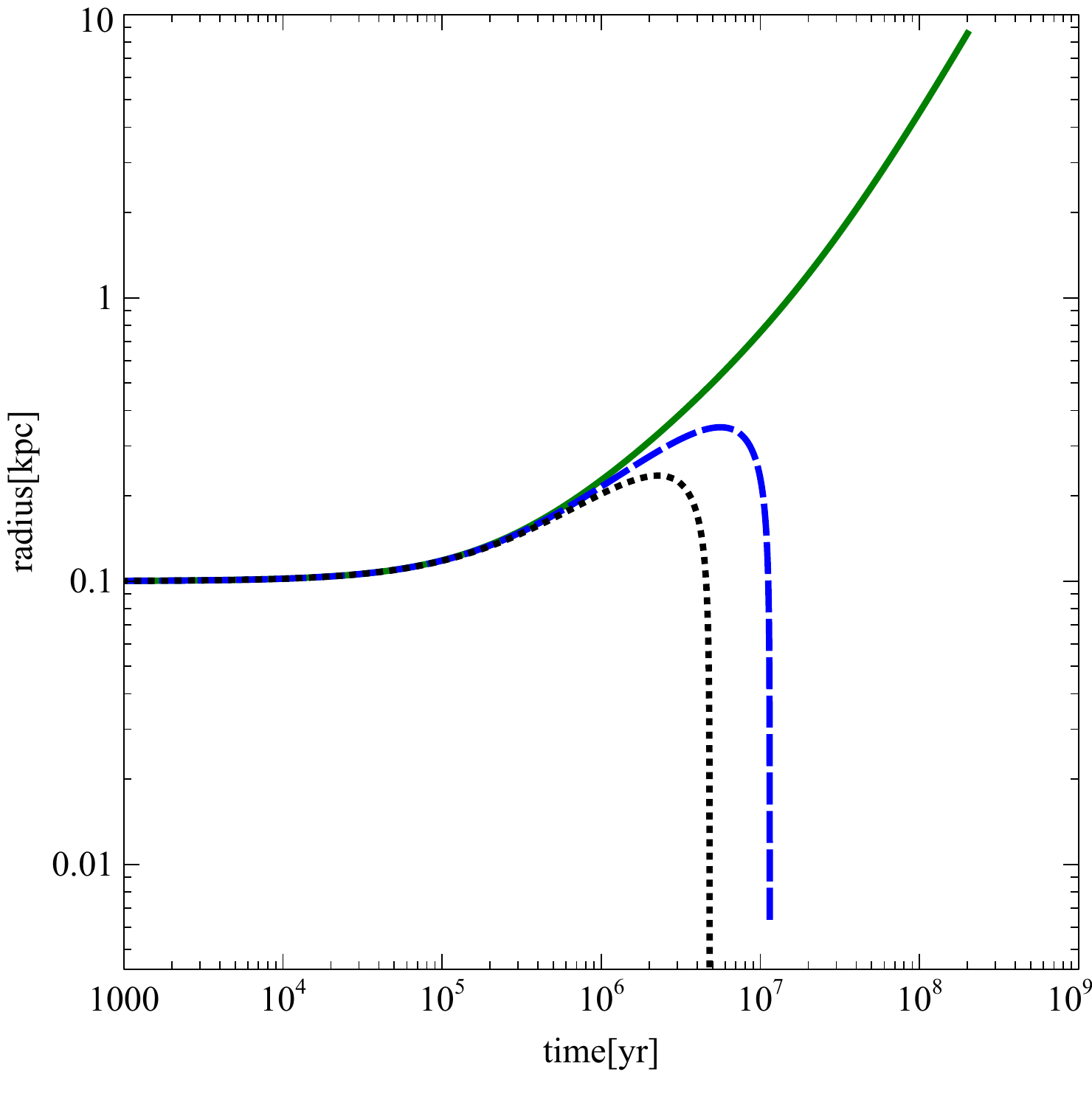} 
\caption{ Evolution of the shell in the isothermal potential for different luminosities, close to the critical value: $L \sim 0.9$ $L_\mathrm{c}$ (black dotted line), $L \sim 0.95$ $L_\mathrm{c}$ (blue dashed line), $L \sim 1.01$ $L_\mathrm{c}$ (green solid line). The initial conditions are the same for the three curves: $r_0$ = 100 pc, $v_0$ = 200  km $\mathrm{s^{-1}}$. The shell can expand outwards only for $L \geq$ $L_\mathrm{c}$. }
\label{Fig_iso_limit}
\end{center}
\end{figure}

\begin{figure}
\begin{center}
\includegraphics[angle=0,width=0.35\textwidth]{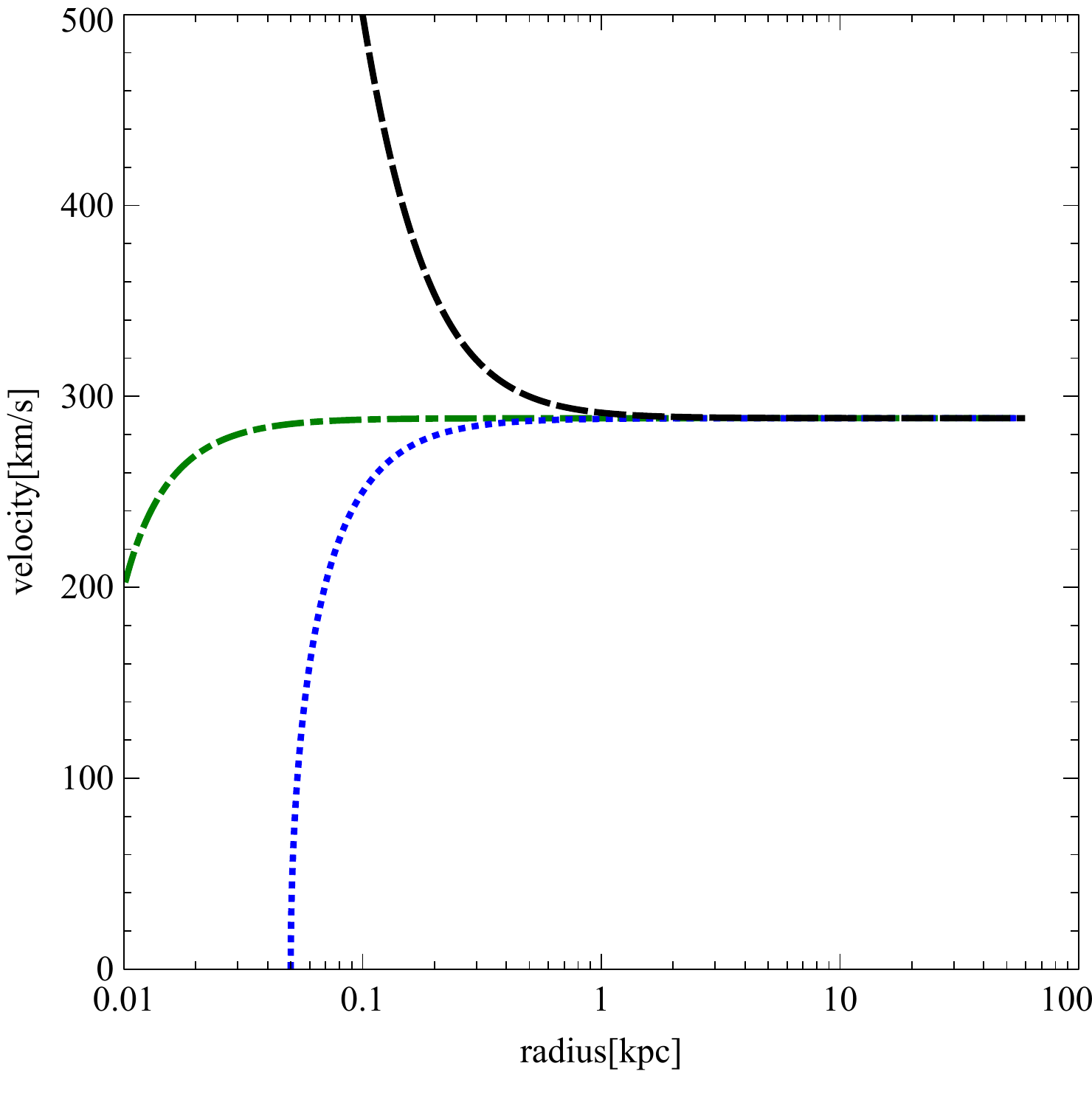}
\caption{Radial velocity profile of the expanding shell in the isothermal potential. The central luminosity is $L \sim 9.4 \times10^{46}$ erg $\mathrm{s^{-1}}$ ($\sim$ 2 $L_\mathrm{c}$); the three curves indicate different initial conditions: $r_0$ = 100 pc, $v_0$ = 500  km $\mathrm{s^{-1}}$ (black dashed line), $r_0$ = 10 pc, $v_0$ = 200  km $\mathrm{s^{-1}}$ (green dash-dot line), $r_0$ = 50 pc, $v_0$ = 0  km $\mathrm{s^{-1}}$ (blue dotted line).
At large radii the shell reaches a constant asymptotic velocity, independent of the initial conditions. }
\label{Fig_iso_velrad}
\end{center}
\end{figure}

The escape or trapping condition of the shell in the galactic potential is mainly determined by a critical luminosity. 
The critical luminosity is defined by considering the equilibrium between outward radiation pressure and inward gravitational forces, which leads to
\begin{equation}
L_\mathrm{c} = \frac{4 f_\mathrm{g} c \sigma^4}{G}  \, . 
\label{L_crit}
\end{equation} 
We see that in the isothermal case the critical luminosity is constant and independent of radius. 
The numerical value is of the order of $L_\mathrm{c} \cong 4.6 \times 10^{46} f_\mathrm{g,0.16} \sigma_{200}^4$  erg $\mathrm{s^{-1}}$ for typical values of the velocity dispersion, $\sigma = \sigma_{200} \times 200$  km $\mathrm{s^{-1}}$, and gas mass fraction, $f_\mathrm{g} = f_\mathrm{g, 0.16} \times 0.16$. 
For luminosities lower than the critical value ($L < L_\mathrm{c}$), the outward force due to radiation pressure is not sufficient to compensate the inward gravitational attraction: the shell falls back, independently of the initial conditions.
But once the central source exceeds the critical luminosity ($L \geq L_\mathrm{c}$) the shell is able to expand outwards indefinitely. The evolution of the shell for different central luminosities is shown in Fig. \ref{Fig_iso_limit}. 
In Fig. \ref{Fig_iso_velrad} we plot the radial velocity profile of the expanding shell for different initial conditions. 
We see that at large radii ($r \gg r_0$) the expanding shell always reaches a constant velocity, independently of the initial conditions.  
From Eq. (\ref{Eq_vr}) we can derive the value of the asymptotic velocity, $v_\mathrm{\infty} = v( r \gg r_0)$: 
\begin{equation}
v_\mathrm{\infty} = \sqrt{ \left( \frac{G L}{2 f_\mathrm{g} c \sigma^2} - 2 \sigma^2 \right) } \, . 
\end{equation} 
We observe that the asymptotic velocity has no dependence on the initial conditions, and is primarily determined by the central luminosity. The shell velocity is larger for higher luminosities, scaling as $v_\mathrm{\infty} \propto \sqrt{L}$.
The velocity of the outflowing shell can be compared with the local escape velocity. 
For an isothermal sphere, the escape velocity is constant and independent of radius: $v_\mathrm{esc} = 2 \sigma$. 
The formal condition for the escape of the shell is given by
\begin{equation}
v_\mathrm{\infty} \geq v_\mathrm{esc} \quad \Rightarrow \quad \sqrt{ \left( \frac{G L}{2 f_\mathrm{g} c \sigma^2} - 2 \sigma^2 \right) } \geq 2 \sigma \, . 
\label{Eq_escape}
\end{equation} 
The escape condition (Eq. \ref{Eq_escape}) then implies that the central luminosity should exceed a critical value: 
\begin{equation}
L \geq L_\mathrm{c}^{\prime} = \frac{12 f_\mathrm{g} c \sigma^4}{G} = 3 L_\mathrm{c} \, .
\end{equation} 
This defines another critical luminosity, which is equal to three times the critical luminosity for the shell expansion derived in Eq. (\ref{L_crit}). 
It implies that the luminosity of the central source should at least reach $L_\mathrm{c}^{\prime}  = 3 L_\mathrm{c}$ for the shell to be able to escape the galactic halo.


\subsubsection{Implications for the $M_\mathrm{BH} - \sigma$ relation}

The luminosity of the central source is related to the mass of the central black hole.  
If one assumes the standard Eddington limit for accretion onto the black hole
\begin{equation}
L_\mathrm{E} = \frac{4 \pi G c m_\mathrm{p}}{\sigma_\mathrm{T}} M_\mathrm{BH} = \frac{4 \pi G c}{\kappa} M_\mathrm{BH} \, , 
\end{equation} 
where $\kappa = \sigma_\mathrm{T}/m_\mathrm{p}$ is the electron scattering opacity, the escape condition (Eq. \ref{Eq_escape}) implies: 
\begin{equation}
M_\mathrm{BH} \geq 3 \frac{f_\mathrm{g} \kappa}{\pi G^2} \sigma^4 = 3 M_\mathrm{\sigma} \, . 
\end{equation} 
Note that $\kappa$ in the above expression refers only to the interaction of radiation and matter limiting accretion onto the black hole, and not to the interaction with the interstellar medium of the host galaxy. 
We thus recover the standard $M_\mathrm{BH} - \sigma$ relation, but with an additional factor of three compared to the usual value of $M_\mathrm{\sigma}$. This suggests that the condition $M_\mathrm{BH} =  M_\mathrm{\sigma}$ only allows a shell to expand towards larger radii, but is not sufficient for the shell to be expelled from the host galaxy. 
In order for the shell to completely escape the galaxy halo, the black hole mass must exceed $3 M_\mathrm{\sigma}$ (see also \citet{McQuillin_McLaughlin_2012}).


\subsection{Hernquist profile}
\label{Subsect_Hernquist}

\begin{figure}
\begin{center}
\includegraphics[angle=0,width=0.35\textwidth]{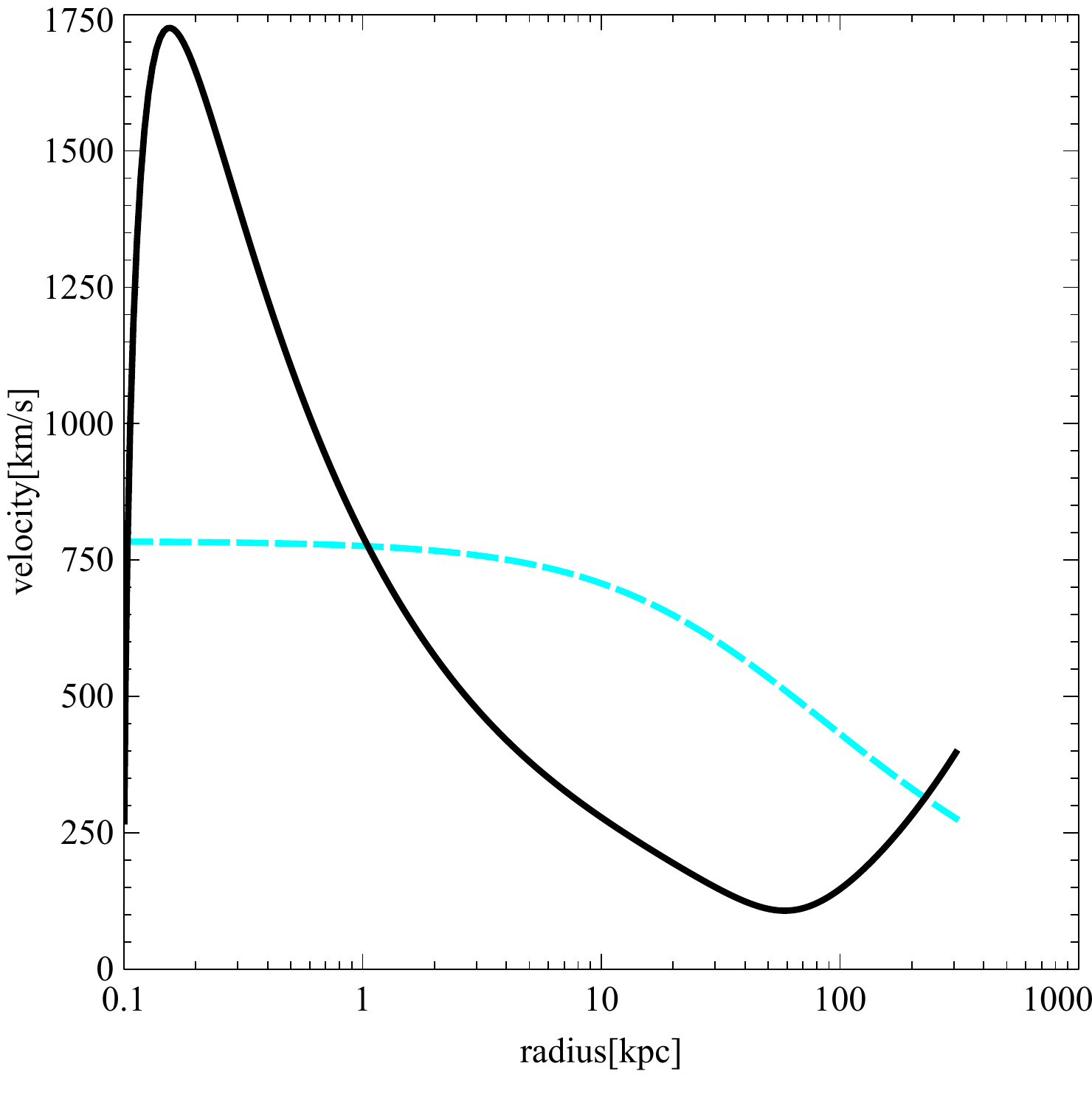} 
\caption{ Radial velocity profile of the expanding shell (black solid line) and local escape speed (cyan dashed line) in a Hernquist potential. The escape velocity is defined as $v_\mathrm{esc} = \sqrt{2 \vert \Phi(r) \vert}$, where the gravitational potential is obtained by integrating the Poisson equation $\nabla^2 \Phi = 4 \pi G \rho$.
The central luminosity is $L_\mathrm{c} \sim 4.6 \times10^{46}$  erg $\mathrm{s^{-1}}$. 
The shell velocity exceeds the local escape velocity at very large radii. }
\label{Fig_Hernquist_Case2_L46e46_v(r)_ve}
\end{center}
\end{figure}

Galaxies are embedded in dark matter haloes, which dominate the outer gravitational potential. 
Dark matter haloes are not ideal isothermal spheres and are better approximated by profiles with varying slopes. 
We first consider the Hernquist density profile \citep{Hernquist_1990} given by 
\begin{equation}
\rho_\mathrm{DM}(r) = \frac{M_\mathrm{DM}}{2 \pi} \frac{r_\mathrm{a}}{r (r+r_\mathrm{a})^3} \, , 
\end{equation}
where $M_\mathrm{DM}$ is the total mass, and $r_\mathrm{a}$ is a characteristic scale length where the density profile changes shape. 
Compared to the isothermal case, the Hernquist density profile is shallower at small radii ($\rho \propto 1/r$) and steeper in the outer regions ($\rho \propto 1/r^4$). 
The corresponding mass distribution is given by
\begin{equation}
M_\mathrm{DM}(r) = M_\mathrm{DM} \frac{r^2}{(r+r_\mathrm{a})^2} \, .
\end{equation}  
The gas mass is taken to be a fraction $f_\mathrm{g}$ of the dark matter mass. 
By analogy with the isothermal case, a critical luminosity can be defined by equating the radiation pressure and the gravitational force terms:
\begin{equation}
L_\mathrm{c,H} = L_c(r_\mathrm{a}) =  c f_\mathrm{g} \frac{G M_\mathrm{DM}^2}{16 r_\mathrm{a}^2} \, . 
\label{Eq_LcH}
\end{equation}
As in the previously discussed isothermal case, once the central luminosity exceeds the critical value, shells can expand outwards to arbitrarily large radii. 
In a Hernquist potential, the shell velocity reaches a minimum around the scale radius and increases at larger radii, eventually exceeding the local escape velocity (Fig. \ref{Fig_Hernquist_Case2_L46e46_v(r)_ve}). 
The shape of the radial velocity profile implies that the shell is able to escape the galactic halo, once the luminosity reaches the critical value given in Eq. (\ref{Eq_LcH}). 
Thus in a Hernquist potential, the critical luminosity required for the shell escape is equal to that defined for the shell expansion. 
However, in order to escape the galactic halo, the shell needs to reach very large radii where its velocity starts to increase and eventually exceed the local escape velocity. 
This implies that AGN feedback should be powerful enough to bring the shell at the required large distances. 

\begin{figure}
\begin{center}
\includegraphics[angle=0,width=0.35\textwidth]{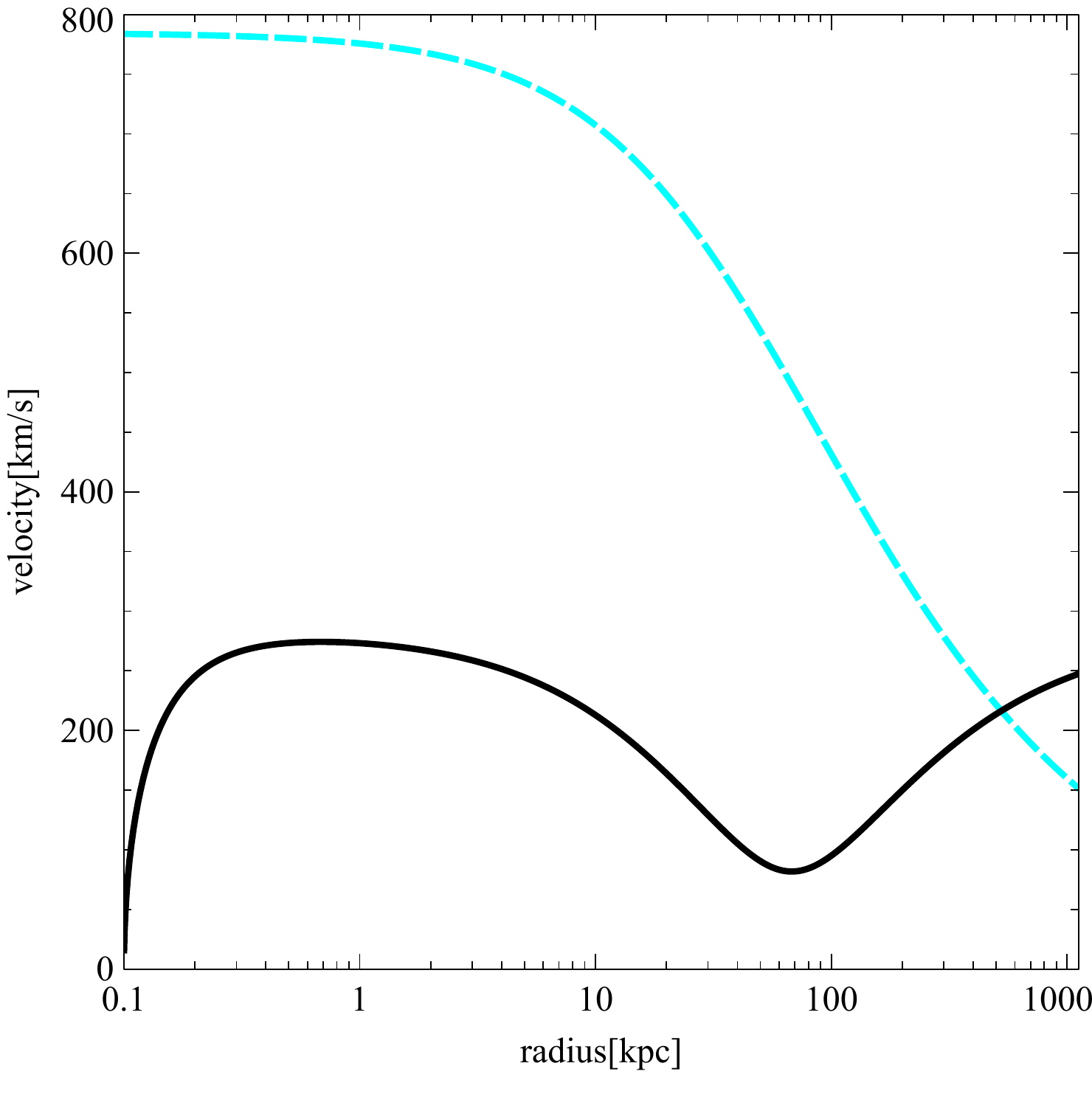} 
\caption{ Radial velocity profile of the expanding shell (black solid line) and local escape speed (cyan dashed line) in a potential composed by isothermal gas and Hernquist dark matter. The central luminosity is $L_\mathrm{c} \sim 4.6 \times10^{46}$  erg $\mathrm{s^{-1}}$. The shell velocity exceeds the local escape velocity at very large radii. }
\label{Fig_iso_H}
\end{center}
\end{figure}

In the above discussion we have implicitly assumed that the cold gas directly traces the dark matter distribution. 
But this assumption may not be adequate, in particular in the inner regions.  
The stellar density distribution generally follows $\rho \propto 1/r^2$, and cold gas from which stars ultimately form should also follow a similar distribution. 
Therefore one may separate the two contributions in the equation of motion, and assume an isothermal distribution for the gas along with a Hernquist profile for the dark matter component. 
The resulting velocity profile is shown in Fig. \ref{Fig_iso_H}. 
We observe that the shell does not accelerate to very high velocities at small radii, and is similar to the isothermal case in the inner regions; while at outer radii the velocity profile follows the characteristic shape of the Hernquist potential, which potentially allows the shell escape. 

\subsection{NFW profile}
\label{Subsect_NFW}

We next consider the so-called NFW profile \citep*{NFW_1996, NFW_1997}, assumed to be a universal profile describing dark matter haloes. 
The density distribution is given by
\begin{equation}
\rho(r) = \frac{\rho_\mathrm{c} \delta_\mathrm{c}}{(r/r_\mathrm{s})(1+r/r_\mathrm{s})^2} \, , 
\end{equation} 
where $\rho_\mathrm{c} = \frac{3 H^2}{8 \pi G}$ is the critical density, $\delta_\mathrm{c}$ is a characteristic density, and $r_\mathrm{s}$ is a scale radius. 
At large radii $(r \gg r_\mathrm{s})$, the NFW profile is shallower ($\rho \propto 1/r^3$) than the Hernquist profile; while both profiles tend to $\rho \propto 1/r$ in the central regions. 
The associated mass profile is given by 
\begin{equation}
M(r) = 4 \pi \delta_\mathrm{c} \rho_\mathrm{c} r_\mathrm{s}^3 \, \left[ \ln \left(1+\frac{r}{r_\mathrm{s}}\right) - \frac{r}{r+r_\mathrm{s}} \right] \, .
\end{equation} 

\begin{figure}
\begin{center}
\includegraphics[angle=0,width=0.35\textwidth]{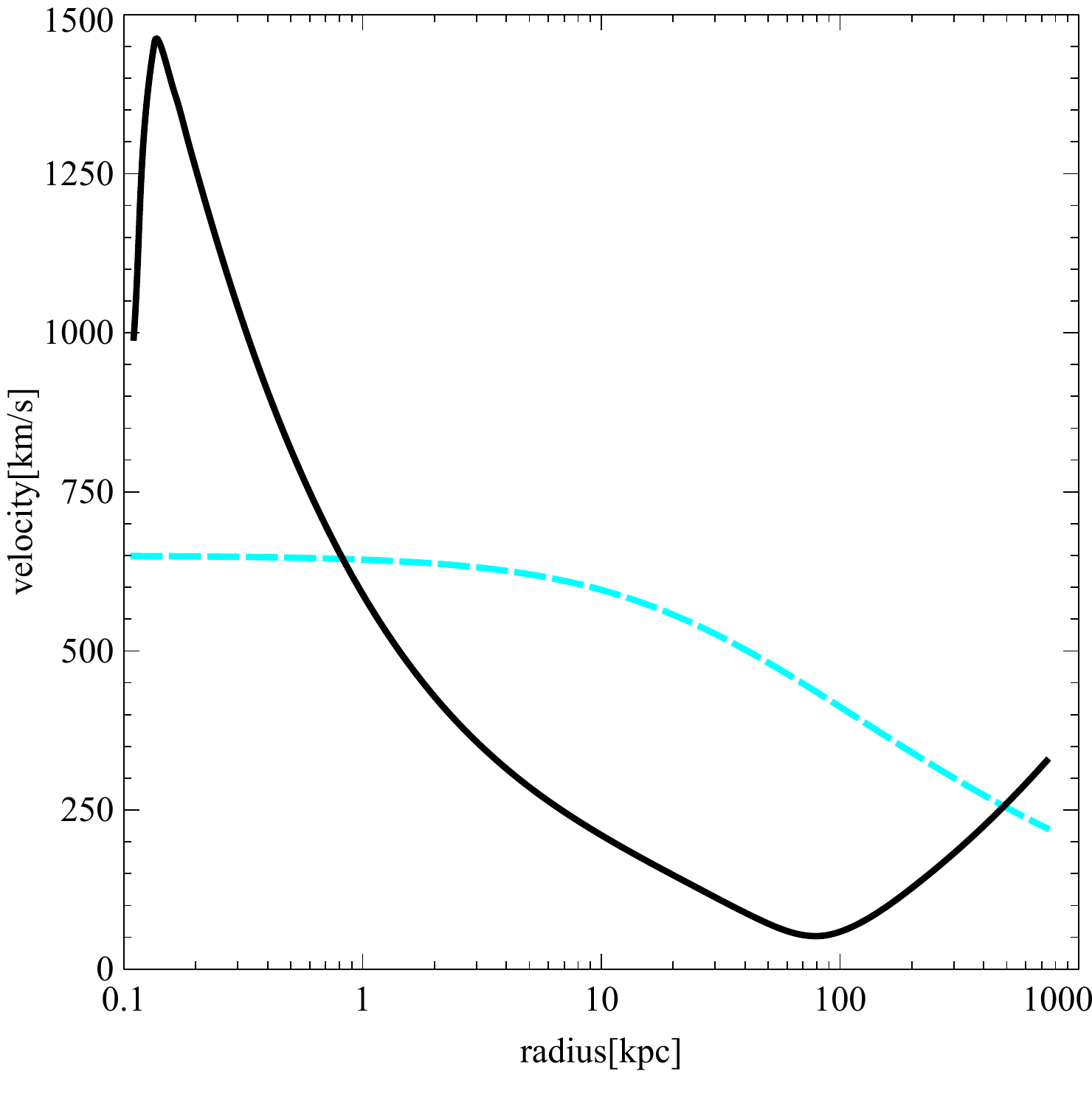} 
\caption{ Radial velocity profile of the expanding shell (black solid line) and local escape speed (cyan dashed line) in a NFW potential. The central luminosity is $L_\mathrm{c} \sim 1.5 \times10^{46}$  erg $\mathrm{s^{-1}}$. 
The shell velocity exceeds the local escape velocity at very large radii.}
\label{Fig_NFW_r(t)_L_limit}
\end{center}
\end{figure}

We repeat the same analysis as for the Hernquist potential case, and first determine the critical luminosity defined as the maximum of the $L_\mathrm{c}(r)$ function. 
As in the previously discussed cases, expanding shells can reach large radii only for luminosities exceeding the critical value. 
The velocity profile of the NFW potential is qualitatively similar to the Hernquist case: the shell velocity reaches a minimum beyond which it starts to increase, eventually exceeding the local escape velocity at very large radii (Fig. \ref{Fig_NFW_r(t)_L_limit}). 
Considering an isothermal gas distribution and a NFW dark matter gives a qualitatively similar result to the corresponding case in the Hernquist potential.

\section{Triggered star formation}
\label{Sect_SF}

In the previous sections we have analysed the evolution of the radiation pressure driven shell in different gravitational potentials. The surrounding medium is swept up by the expanding shell; the squeezing and compression of the cold gas induced by the passage of the shell may cause local density enhancements.
This could, in turn, lead to a triggering of star formation within the outflowing shell.
We adopt a simple prescription for the triggered star formation in the shell and estimate the star formation rate as:
\begin{equation}
\dot M_{\star} \sim \epsilon_{\star} \,  \frac{M_\mathrm{g}(r)}{t_\mathrm{flow}(r)} \, , 
\label{Eq_SFR}
\end{equation}
where $\epsilon_{\star}$ is the star formation efficiency, and $t_\mathrm{flow}(r) = \frac{r}{v(r)}$ is the local flow time. 
The observed star formation efficiency is typically of the order of a few percent, and somewhat higher in starburst systems. 
In the isothermal case, the gas mass scales with radius and Eq. (\ref{Eq_SFR}) leads to
\begin{equation}
\dot M_{\star} \sim \epsilon_{\star} \,  \frac{2 f_\mathrm{g} \sigma^2}{G} v(r) \, . 
\end{equation}
We observe that the star formation rate directly scales with the velocity of the shell, $\dot M_{\star} \propto v(r)$. 
\begin{figure}
\begin{center}
\includegraphics[angle=0,width=0.4\textwidth]{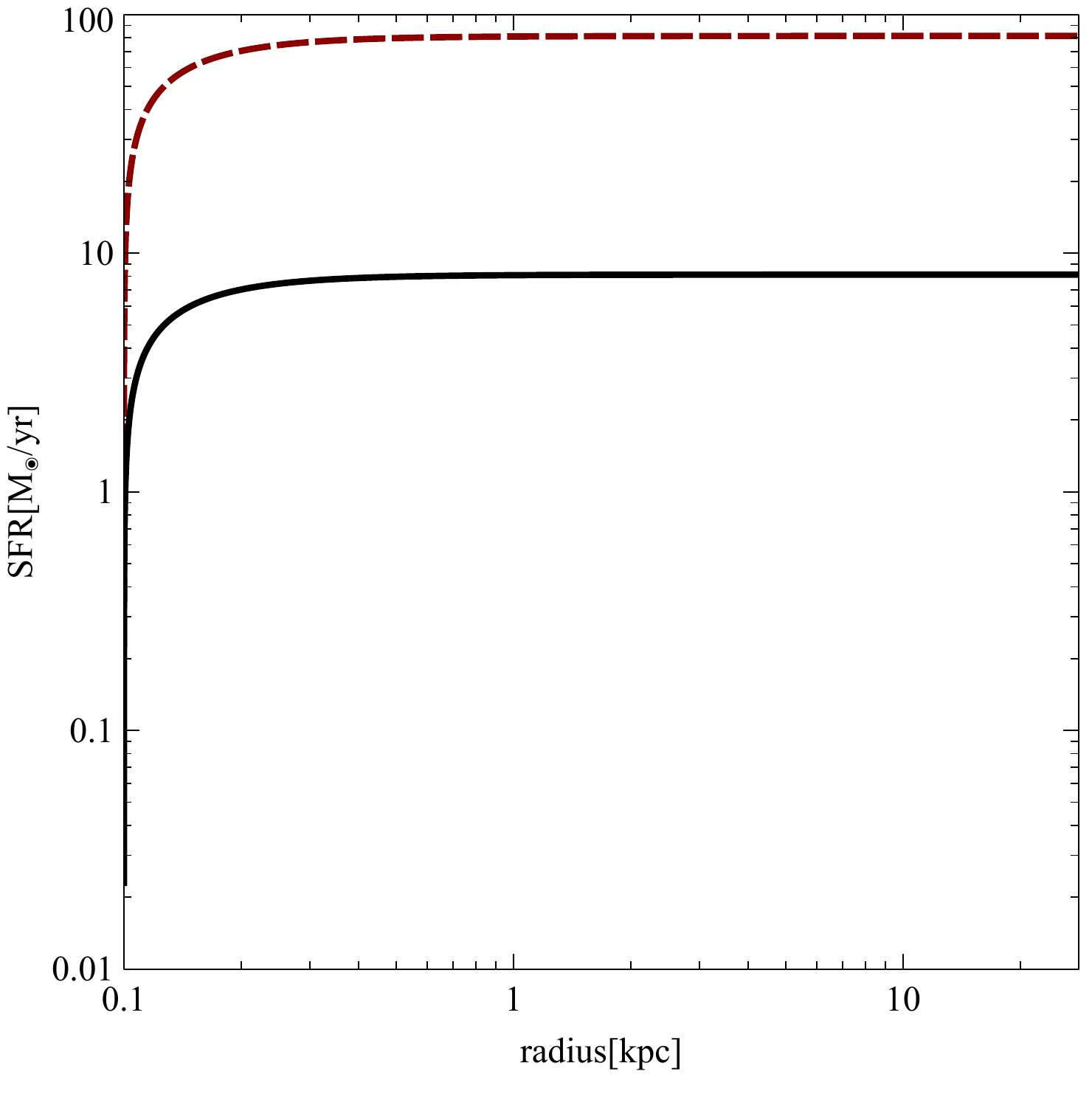}
\caption{ Star formation rate as a function of radius in the isothermal potential. The central luminosity is $L \sim 9.2 \times10^{46}$  erg $\mathrm{s^{-1}}$  ($\sim$ 2 $L_\mathrm{c}$); the two curves indicate different star formation efficiencies: $\epsilon_{\star} = 0.01$ (black solid line), $\epsilon_{\star} = 0.1$ (red dashed line). The star formation rate is constant at large radii. }
\label{Fig_SFR_iso}
\end{center}
\end{figure}
We have seen that in the isothermal case the shell velocity tends towards a constant asymptotic value at large radii. 
This implies that the star formation rate is also constant at large radii:
\begin{equation}
\dot M_\mathrm{\star, \infty} = \dot M_{\star} (r \rightarrow \infty) \sim \epsilon_{\star} \,  \frac{2 f_\mathrm{g} \sigma^2}{G} v_\mathrm{\infty} \, . 
\end{equation}
As $v_\mathrm{\infty} \propto \sqrt{L} \propto \sqrt{M}$, more luminous or equivalently more massive objects tend to have higher star formation rates. 
For typical galaxy parameters and shell velocities, we obtain star formation rates of the order of $\dot M_\mathrm{\star, \infty} \sim 10-100$ $M_{\odot}$ $\mathrm{yr}^{-1}$, for $\epsilon_{\star} \sim0.01-0.1$, respectively (Fig. \ref{Fig_SFR_iso}). 
Once formed, stars decouple from the expanding shell; the initial velocity of the newly formed stars corresponds to the shell velocity at the formation radius. As long as the initial speed is lower than the local escape velocity, stars that are formed will remain bound to the galaxy.  
Thus star formation in the outflowing shell contributes to the increase of the host stellar mass in the outer regions. 
Integrating over the total AGN feedback lifetime, one may obtain a rough order-of-magnitude estimate for the total additional stellar mass formed in this way. 
In the isothermal case, the total stellar mass can be of the order of several $10^{9}$ $M_{\odot}$ for one AGN episode lasting a Salpeter time ($\sim 5 \times 10^7$ yr). 
\begin{figure}
\begin{center}
\includegraphics[angle=0,width=0.35\textwidth]{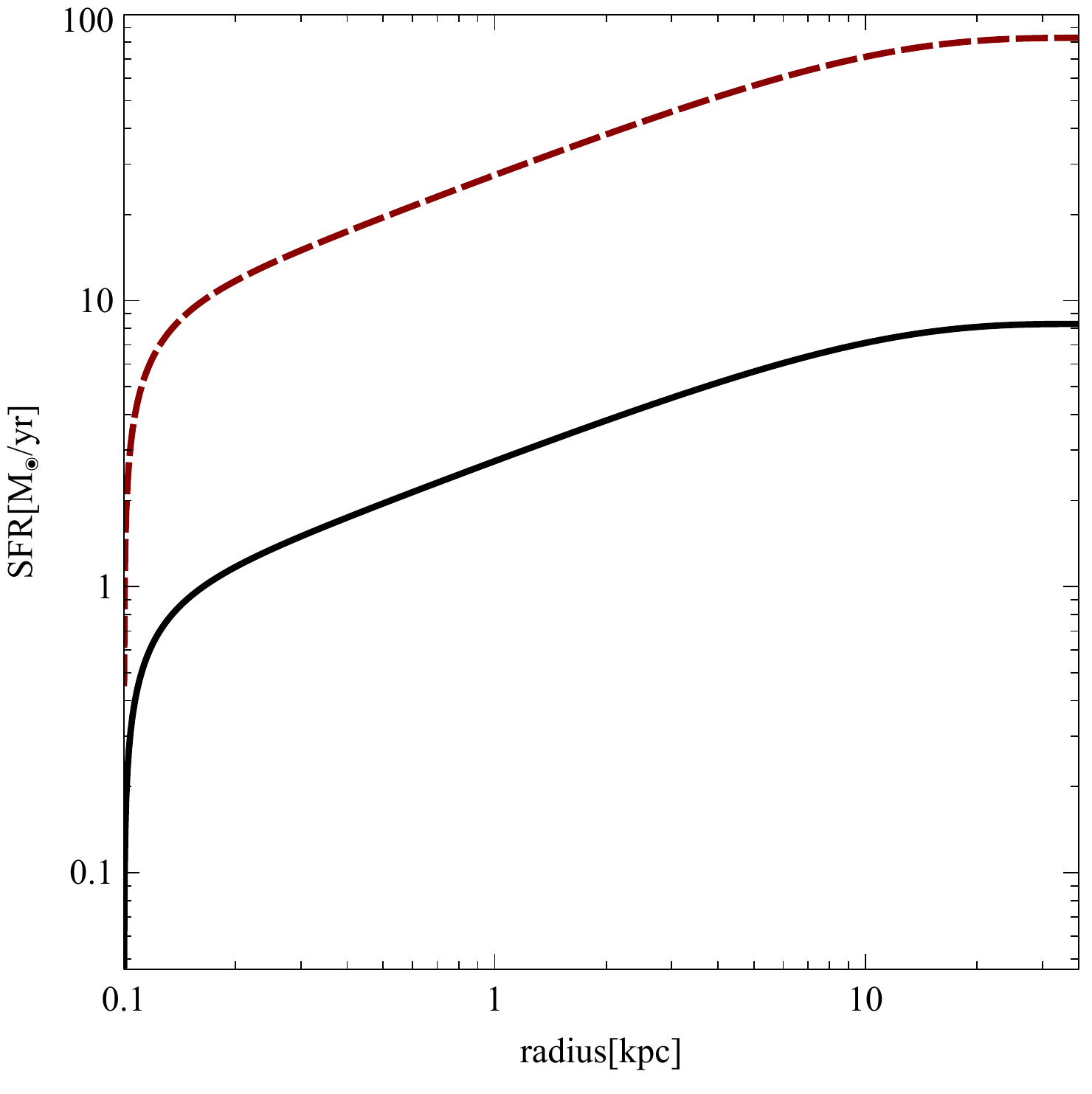} 
\caption{ Star formation rate as a function of radius in the Hernquist potential. The central luminosity is $L \sim 9.2 \times10^{46}$  erg $\mathrm{s^{-1}}$  ($\sim$ 2 $L_\mathrm{c}$), and the line symbols are the same as in Fig. \ref{Fig_SFR_iso}. }
\label{Fig_SFR_H}
\end{center}
\end{figure}
\begin{figure}
\begin{center}
\includegraphics[angle=0,width=0.35\textwidth]{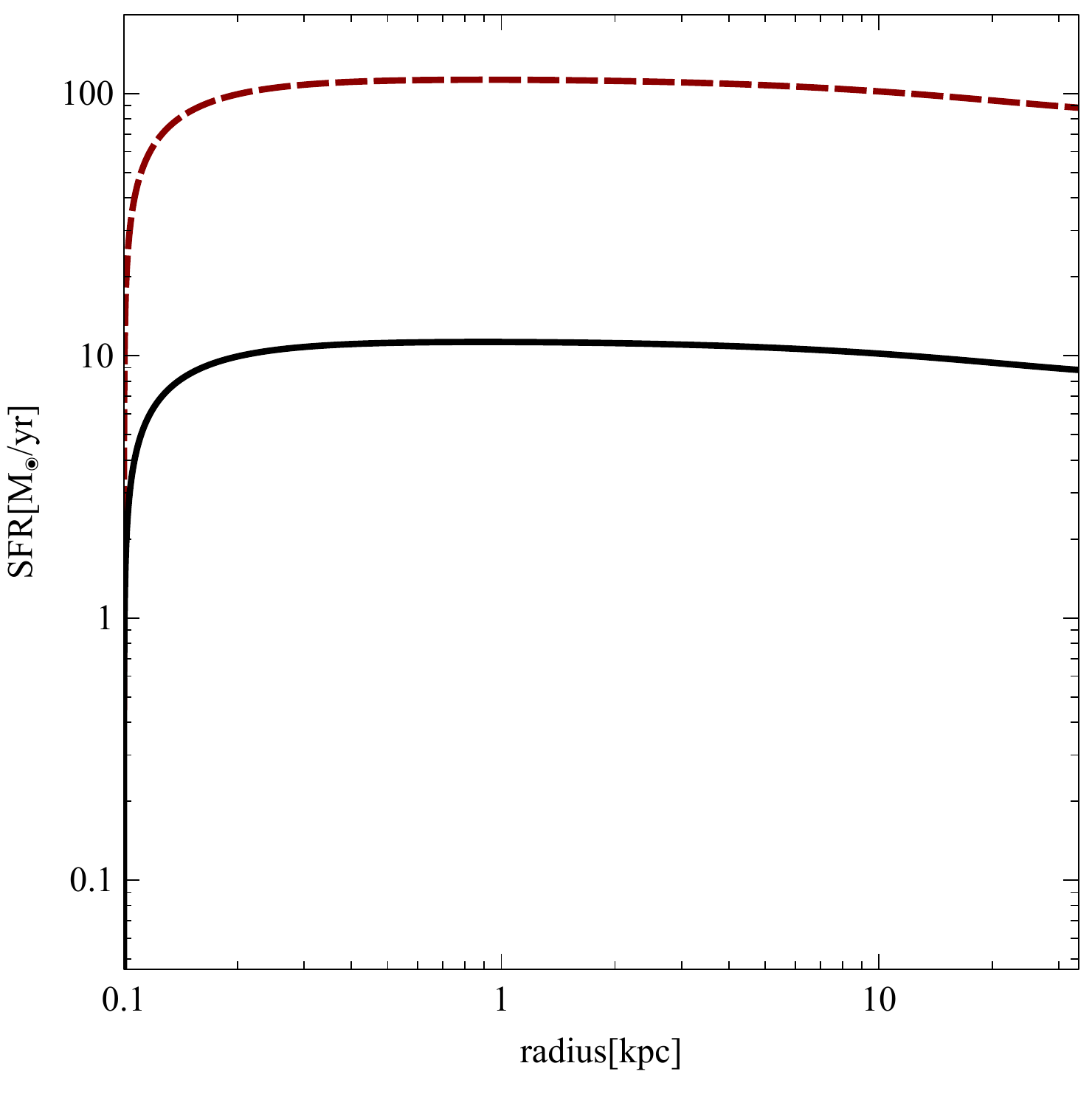}
\caption{ Star formation rate as a function of radius for an isothermal gas distribution and Hernquist dark matter profile. 
The line symbols are the same as in Fig. \ref{Fig_SFR_iso}. }
\label{Fig_SFR_iso_H}
\end{center}
\end{figure} 

The same computations can be performed in the case of the Hernquist potential. 
The resulting star formation rate is shown in Fig. (\ref{Fig_SFR_H}); the associated stellar mass is comparable to the isothermal case. 
For the combined case of isothermal gas and Hernquist dark matter, the radial profile of the star formation rate in the inner regions is very similar to the isothermal case (Fig. \ref{Fig_SFR_iso_H}), and the total additional stellar mass is also of the same order. 
The star formation rate in the case of a NFW potential seems to be somewhat lower, with correspondingly lower stellar masses. 
The combined case of isothermal gas and NFW dark matter is again similar to the corresponding case for the Hernquist potential.


\section{Discussion}
\label{Sect_Discussion}

It is usually assumed that when the central black hole reaches the critical mass, $M_\mathrm{\sigma}$, the resulting feedback blows away the surrounding material, shutting off further black hole growth and suppressing star formation. 
However, we have seen that this condition is not sufficient to completely expel gas from the host galaxy in the case of an isothermal potential. 
In fact, for $M_\mathrm{BH} = M_\mathrm{\sigma}$, the outflowing shell velocity tends to a constant asymptotic value at large radii, which is lower than the local escape velocity.
The shell thus remains trapped in the galactic halo, and once the AGN activity turns off, it may eventually fall back (providing potential fuel for further accretion and star formation). 
The actual condition for the shell escape, i.e. $v_\mathrm{\infty} \geq 2 \sigma$, requires the black hole mass to exceed $M_\mathrm{BH} = 3 M_\mathrm{\sigma}$.  
This is three times the critical value of the standard $M_\mathrm{BH} - \sigma$ relation. 
It has been argued that black holes lying on the observational $M_\mathrm{BH}-\sigma$ relation cannot expel material from the galactic halo by radiation pressure alone, and that an additional momentum boost is required \citep{Silk_Nusser_2010}. 
The shortfall seems to be a factor of a few, which is consistent with our result.  
Recent numerical simulations of radiation pressure-driven feedback also show that for $L \sim L_\mathrm{c}$, gas is only evacuated from the central nuclear region, while the critical luminosity needs to be exceeded by a factor of several in order to clear gas out of the entire galaxy \citep{DeBuhr_et_2010, DeBuhr_et_2011}.
However, if the galaxy is residing in a group or cluster environment, gas in the outer regions could be removed by ram pressure or tidal stripping, and thus environmental effects also play an important role. 
On the other hand, if an initially momentum-driven wind switches to an energy-driven form, then a black hole mass exceeding the $M_\mathrm{BH}-\sigma$ value may not be required \citep{King_et_2011}. 

Previous studies of AGN feedback generally assumed an isothermal profile for both the gas and the dark matter distributions \citep{Silk_Rees_1998, Fabian_1999, Fabian_et_2002, King_2003, King_2005, Silk_Nusser_2010}. 
Recently, momentum-driven feedback in non-isothermal dark matter haloes has also been analysed \citep{McQuillin_McLaughlin_2012}.
Here we follow the evolution of the feedback-driven shell and determine its escape/trapping conditions in different underlying dark matter potentials.  
At large radii, the Hernquist and NFW density profiles are steeper than the isothermal one, and the resulting radial velocity profiles are also qualitatively different: the shell velocity increases and eventually exceeds the local escape velocity at very large radii. 
Therefore escape is formally possible for $M = M_\mathrm{\sigma}$ in a Hernquist or NFW potential, provided that the shell is able to reach such great distances. 
But this in turn requires extremely long timescales for the feedback process, and one should take into account the AGN lifetime. 
For instance in the case of a Hernquist potential, for $M = M_\mathrm{\sigma}$ the shell would take $\sim 10^9$ yr to reach the critical radius (where its velocity becomes larger than the local escape velocity), which is very long compared to typical AGN activity time scales. 
For higher central luminosities, the critical radius may be reached on more plausible time scales of the order of a few Salpeter times ($\sim 10^8$ yr).

Feedback from the central black hole is usually invoked in galaxy formation and evolution models to suppress star formation. 
However AGN feedback may also operate in the opposite direction and even trigger star formation in the host galaxy. 
In fact, triggering of star formation associated with radio jet activity, has been proposed in the past to explain the observed alignment of radio and optical structures in high-redshift radio galaxies \citep{Rees_1989, Begelman_Cioffi_1989}.  
A positive feedback mechanism, induced by the AGN jet and leading to enhanced star formation, has also been considered as a source powering luminous starbursts \citep{Silk_2005}. 
Numerical simulations of radio jet-induced star formation have also been recently performed \citep{Gaibler_et_2012}. 

Here we explicitly consider triggered star formation in the radiation pressure-driven shell, where the characteristic time scale is given by the local flow time $t_\mathrm{flow}(r)$. 
We recall that we are considering cold, dusty, and primarily molecular gas from which stars are expected to form. 
As the shell moves on, stars are gradually dropped out of the expanding shell, and subsequently follow their own orbits determined by the gravitational potential of the galaxy. 
The newly formed stars are expected to follow nearly radial orbits, although the exact stellar orbit depends on the initial conditions, i.e. formation radius and local shell velocity.  
Observations of bright elliptical galaxies show that the velocity dispersion profile steeply declines towards the exterior \citep[][and references therein]{Romanowsky_et_2003, Gerhard_2010}. This indicates that the orbits of the stars in the outer halos of bright ellipticals are dominated by radial motion, which may be consistent with our interpretation. 
Overall, star formation in the outflowing shell suggests that new stars are formed at increasingly larger radii, successively populating the outer regions of the host galaxy. 

This characteristic pattern may be compared with observational studies of the outer growth of galaxies. 
Detailed observations of the growth of massive galaxies from redshift $z \sim 2$ to the present have shown that the total stellar mass of these galaxies has increased by a factor of $\sim$2 over this time range \citep{vanDokkum_et_2010}. 
But this doubling does not occur in a homogeneous way throughout the galaxy: the stellar mass in the central regions remains roughly constant, while the increase in mass mainly occurs at outer radii. 
This suggests that the mass growth takes place in the outer regions of the galaxy with a gradual build-up of the stellar mass at larger radii, implying a sort of `inside-out' growth. 
Comparison of a sample of compact, quiescent galaxies at $z \sim 2.3$ with nearby massive elliptical galaxies suggests that the compact high-redshift objects may form the central cores of today's ellipticals \citep{Bezanson_et_2009}.  
Recent measurements also confirm the extremely small sizes of massive quiescent galaxies at high redshift \citep{Szomoru_et_2012}.
All these observations seem to support the inside-out growth scenario, and point toward an evolutionary scenario whereby the galaxy growth mainly takes place in the outer regions, possibly by accretion of stars through merger processes. 

Alternatively, the observed outer growth of massive galaxies may be interpreted in terms of AGN feedback, and the above sketched picture of star formation in the feedback-driven shell seems at least qualitatively consistent with observations. 
However, we should mention that observational studies usually analyse the quiescent phases of galaxies in selected samples at given redshifts; whereas we discuss active phases when the central AGN is switched on and is triggering star formation. Observationally such activity phases may be better classified as quasars or ultra-luminous infrared galaxies. 
One can consider several episodes of AGN activity and associated feedback.  Accretion of material, via some form of cold flows, should replenish the gas content of the host galaxy. 
Gas added to the galaxy will then fuel both the central black hole and feedback-driven star formation; the black hole mass increases and so does the stellar velocity dispersion. As a result of the repetition of many such episodes, galaxies will stay on the $M_\mathrm{BH} - \sigma$ relation. 

It is also interesting to note that recent observations indicate the existence of massive galaxies at $z \sim 2$ which are much more compact compared to galaxies of comparable mass at $z \sim 0$. 
These findings imply a significant evolution in size, with an increase of a factor of $\sim 5$ in effective radius, but without a correspondingly strong increase in mass. 
Different physical interpretations of the size evolution of galaxies have been considered, including major mergers, minor mergers, and adiabatic expansion \citep{Naab_et_2009, Bezanson_et_2009, Fan_et_2008}. 
The observed size evolution may also be interpreted in the framework of the AGN feedback-driven star formation. 
The next step will be to analyse more in detail the fate of the newly formed stars by computing their subsequent orbits in the gravitational potential of the host galaxy. This should allow us to make more quantitative comparisons with the observed `inside-out' growth of massive galaxies. 
A succession of AGN feedback episodes should also be considered and its potential role in the build-up and evolution of the host galaxy.

\section*{Acknowledgments}

We thank Becky Canning for interesting discussions. 
WI acknowledges support from the Swiss National Science Foundation.

\bibliographystyle{mn2e}
\bibliography{biblio.bib}


\label{lastpage}

\end{document}